\documentclass[useAMS,usenatbib, usegraphicx]{mn2e}

\include{epsf}

\def\apj{{\rm ApJ}}
\def\mnras{{\rm MNRAS}}

\def\etal{{\rm et~al.\ }}
\def\hmpc{\;h^{-1}{\rm Mpc}}

\def\invhmpccc{\;h^{-3}{\rm Mpc}^{3}}

\def\kms{{\rm \;km\;s^{-1}}}

\def\msun{{\rm M_{\odot}}}
\def\hmsun{\;h^{-1}{\rm M_{\odot}}}

\def\simlt{\lower.5ex\hbox{$\; \buildrel < \over \sim \;$}}
\def\simgt{\lower.5ex\hbox{$\; \buildrel > \over \sim \;$}}

\title[Real space Alcock-Paczy\'{n}ski test]
{A potentially pure test of cosmic geometry: galaxy clusters and the
real space Alcock-Paczy\'{n}ski test}
\author[Y.-R. Kim and R. A. C. Croft ]{Young-Rae Kim$^{1}$\thanks{E-mail:
    yr@cmu.edu} and Rupert A. C. Croft$^{1}$\\
$^{1}$Physics Department, 
Carnegie Mellon University, Pittsburgh, PA 15213, USA}
\begin{document}

\date{\today}

\pagerange{\pageref{firstpage}--\pageref{lastpage}} \pubyear{2005}

\maketitle
\label{firstpage}

\begin{abstract}
We investigate the possibility of probing dark energy by measuring 
the isotropy of the galaxy cluster autocorrelation function 
(an Alcock-Paczy\'{n}ski test). The correlation function
is distorted in redshift space because of the cluster peculiar
velocities, but if these are known and can be
subtracted, the correlation function 
measurement becomes in principle a pure test of cosmic geometry.
Galaxy cluster peculiar velocities can be measured using the
kinetic Sunyaev Zel'dovich (kSZ) effect. Upcoming CMB surveys, e.g., 
ACT, SPT, Planck, are expected to do this with
varying levels of accuracy, dependent on systematic errors 
due to cluster temperature measurements, radio point
sources, and the primary CMB anisotropy. 
We use the Hubble volume $N$-body simulation and
the  hydrodynamic simulation results of Nagai et. al (2003) to simulate
various kSZ surveys. We find by model fitting that
a measurement of the correlation function distortion can be used to 
recover the cosmological parameters that have been used to generate the
simulation. However, the low space density of 
galaxy clusters requires larger surveys than are taking place
at present to place tight constraints on cosmology.
For example, with the SPT and ACT surveys, $\Omega_{\Lambda}$ could be
measured to within 0.1 and 0.2 respectively at one sigma,
 but only upper limits
on the equation of state parameter $w$ will be possible.
 Nevertheless, with accurate measurements of the kSZ
effect, this test can eventually be used to probe the dark energy
equation of state and its evolution with redshift, with different
systematic errors than other methods.
\end{abstract}


\section{Introduction}

The Alcock-Paczy\'{n}ski (1979, hereafter AP) 
test is a conceptually simple probe of
cosmic geometry and hence dark
energy. It is based on the fact that an 
intrinsically isotropic object will appear distorted
when we translate it from observed space (angles and redshifts)  
into real space (units of length) if we use the wrong cosmology 
(to relate angles and redshift to sizes). The strength of the
distortion will depend on cosmic geometry, for example
the value of the cosmological constant $\Lambda$. The test is 
a way to measure dark energy which requires only that we
find and measure known perfectly symmetric objects in
the Universe. In the present work
 we use the autocorrelation function of galaxy clusters as
the ``isotropic'' object. In addition to the possible geometric distortions,
the correlation function will have a distortion of
different kind, which is due to peculiar velocities of clusters.
This affects the line of sight component of cluster pair
separations (e.g., as applied to galaxies by Davis and Peebles 1983,
 Kaiser 1987, and clusters by 
the Padilla \& Baugh 2002). There is a way
to remove this distortion however, by making use of the 
 kinetic Sunyaev-Zeldovich (kSZ) effect (Sunyaev \& Zeldovich 1972). The kSZ 
effect is the Doppler shift of CMB photons due to
a hot intracluster medium. The CMB
photons interact with free electrons and become Doppler shifted when 
the cluster is moving with respect to 
the CMB rest frame. This causes temperature fluctuations that depend on the
bulk velocity of cluster. There are currently 
on-going and future surveys that 
plan to measure this effect (see e.g., Vale and White 2005 for strategies). 
At the time of writing, the kSZ effect has not yet been detected
for single clusters, although upper limits on 
have been estimated on the
bulk velocity averaged over several clusters (e.g., 1420 $\kms$ at 95 $\%$
confidence by Benson \etal 2003).
 In the present paper we investigate the correction of
 redshift distortions from peculiar velocities 
using kSZ measurements, and how this can leave the pure geometric
distortions from which we can find cosmological parameters. 

Similar methods have been put forward for constraining
cosmology with the clustering of different objects. Phillipps (1994)
suggested using quasar clustering for the Alcock Paczy\'{n}ski test
(averaging over many close pairs of quasars). Several other
studies have adopted the quasar correlation function 
as the isotropic object (e.g., Popowski \etal 1998, Hoyle
\etal 2002, da \^{A}ngela \etal 2005). The difficulty of this 
method is that we only observe the positions of quasars
in redshift space, where the
correlation function is distorted by peculiar velocities (e.g., Kaiser 1987,
Hamilton 1992) and by other source of noise, such redshift errors.   

Many studies have suggested using the 
kinetic Sunyaev-Zel'dovich effect (e.g., Haehnelt
\& Tegmark 1996, Lange \etal 1998, Kashilinsky \& Atrio-Barandela 2000, Aghanim
\etal 2001, Atrio-Barandela \etal 2004, etc.) to measure peculiar velocities
of galaxy clusters.
There are several upcoming sky surveys that aim to do this, such 
as ACT\footnote{http://www.hep.upenn.edu/~angelica/act/act.html},
 SPT\footnote{http://spt.uchicago.edu/} and
Planck\footnote{http://www.rssd.esa.int/Planck/}
 In principle, these types of measurements  
 can  be used to eliminate the distortion due to peculiar 
velocities and therefore we can recover the
isotropic correlation function. The bulk cluster peculiar
velocity is not the sole factor in the 
distortion and there are sources of
noise in kSZ measurements. 
These include microwave background
fluctuations,  cluster internal velocities, and point source contamination
 (see simulations by Haehnelt \& Tegmark 1996, Aghanim \etal 2005.)

In the present work, we use large N-body simulations (the Hubble volume,
Evrard \etal 2002), and the hydrodynamic simulation results of Nagai \etal
(2003)
 to test how well the peculiar
velocity-corrected galaxy cluster AP test can be carried out.
Our plan for the paper is as follows:
In \S2 we discuss the theoretical cluster correlation function we
use as a model as well as that measured from the Hubble volume
simulation. In \S3.1, we explain the model fitting procedure and predict the
effect of kSZ measurements. We describe and measure the geometric
distortions as a function of redshift.
Using the recovered values we constrain cosmological parameters 
from the simulation data, discussing the results in the same
section. We show results from simulations of particular
surveys (SPT and ACT) in  \S4.1-\S4.2. 
We summarize our results and conclude in \S5. 

\section{Cluster Correlation Function}

The cluster autocorrelation function $\xi(r)$ is a measure of 
the probability of
finding a cluster at a distance $r$ from another cluster;
typically it is computed as the ratio of the number of clusters at
a distance $r$ from 
another cluster [$DD(r)$], divided by the number expected in the absence of
clustering [$DR(r)$] minus 1, i.e. $\xi(r) = \frac{DD(r)}{DR(r)} - 1 $,
where DD represent cluster-cluster pairs and DR cluster-random pairs.
$\xi(\sigma, \pi)$ carries
the same definition but the distance between two clusters is represented
in terms of the component directions perpendicular ($\sigma$)
and parallel to
the line of sight ($\pi$). 
Many groups have studied clustering of 
galaxies, clusters and QSOs or the cross-correlation of one 
with another using
simulations and observations (e.g., Mo \& White 1996, Mo
\etal 1996, Borgani \etal 1997, Croft \etal 1999, Colberg \etal 2000,
Moscardini \etal, 2000, 
Zehavi \etal 2002, Hawkins \etal 2003, 
Croom \etal, 2005, Springel \etal, 2005). Springel
\etal (2005)  compare the galaxy 2 point correlation function from their 
high-resolution simulation with 
that of galaxies in the 2dF, SDSS and APM 
surveys and show that the correlation functions
follow power laws for $r<\sim 20 \hmpc$ and that
observation and simulation
agree very well.

 In the most recent Sloan Digital
Sky Survey (SDSS) analysis, Zehavi \etal (2005) studied
the luminosity and color dependence of the
 galaxy correlation function. It was found that
clustering of blue galaxies increases continuously with luminosity, whereas
bright red galaxies show strong clustering at large scales and faint ones
mainly cluster at small scales. Many of the studies described 
fit the correlation function with a single power-law
although this simple form was found be some not to be good enough to use 
(e.g., in analysis of the 2dF data by Hawkins \etal 2003, Croom \etal,
 2005, da \^{A}ngela \etal 2005). These considerations will affect our
choice of model correlation function, described below.

\subsection{Theory}
Alcock \& Paczy\'{n}ski (1979) showed that the ratio of
angular and redshift sizes of a spherical comoving object
 evolve differently in time  
for different cosmologies. Detailed discussions of  how the
components evolve
in redshift are presented in Ballinger \etal (1996) and Matsubara
and Suto (1996).

\begin{figure}
\begin{center}
\epsfxsize=8.0cm
\epsfbox{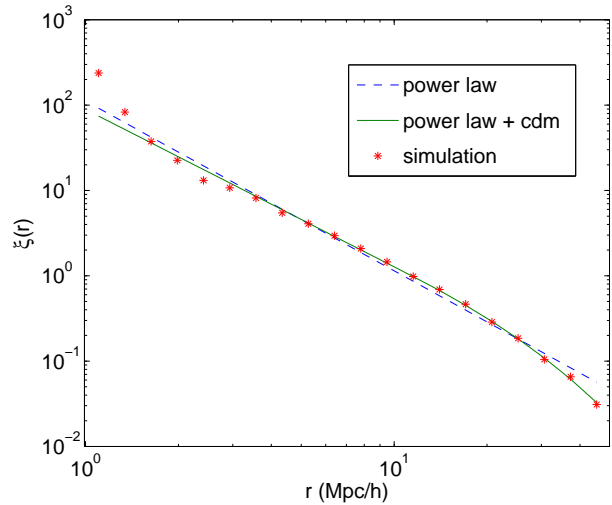}
\caption{The galaxy cluster correlation function from an octant of the Hubble
volume simulation (points), as well as the best fit power law and our
chosen fitting function (power law + linear CDM, see text.)}
\label{fig:adj}
\end{center}
\end{figure}

 Da \^{A}ngela \etal
(2005) argued that the correlation function (of high-z galaxies) cannot
be  described adequately with a single power law, and 
use a double power law when they
fit it. We adopt this general
idea when we build our model
correlation function but instead of using two
power laws, we use a single power-law for $r \leq r_{0}$ and
 use a correlation
function computed using cold dark matter linear theory (the
transfer function from Bardeen \etal 1986)
 for $r>r_{0}$. We test this idea using the
 $\Lambda$CDM Hubble volume simulation 
(see \S2.2 for more detailed description.)
Figure~\ref{fig:adj} shows the best-fit power law 
and the best-fit composite power-law and CDM
 correlation functions. It demonstrates that the latter describes 
the correlation
 function better, especially for large $r$ where the $\xi(r)$
 decreases faster than a power law.

A power law correlation function is given by:
\begin{equation}
\xi(r) = \left(\frac{r}{r_{0}} \right)^{-\gamma}.
\label{eqn:xit_r}
\end{equation}
$r$ can be rewritten in terms of line of sight and transverse components,
\begin{eqnarray}
%r^{2} & = & r_{\perp}^{2} + r_{\parallel}^{2} \nonumber \\
r^{2} & = & \sigma^{2} + \pi^{2} \nonumber \\
      & = & f^{2} \Delta\theta^{2} + g^{2} \Delta z^{2}
\label{eqn:r_decomp}
\end{eqnarray}
where $f$ and $g$ are given by 
\begin{equation}
f = (1+z)D_{A}(z),
\end{equation}
and
\begin{equation}
g = \frac {c}{H(z)}.
\end{equation}
In order to incorporate the  equation of state
of dark energy, we use (Seo \& Eisenstein 2003)
\begin{equation}
H(z) = H_{0} \sqrt{\Omega_{m}(1+z)^{3}+\Omega_{X}  \exp \left[3\int_{0}^{z}
      \frac{1+w(z)}{1+z}dz \right]}
\label{eqn:seo}
\end{equation}
and
\begin{equation}
D_{A}(z) = \frac{c}{1+z} \int_{0}^{z}\frac{dz}{H(z)}.
\end{equation}

In our study, we assume a flat universe such that $\Omega_{m} + \Omega_{X} = 1$ and allow a redshift dependent equation of state with the simple
 form. $w = w_{0} +
w_{1}z$ (e.g., Seo \& Eisenstein 2003).
We introduce a distortion parameter $h$ (Popowski \etal 1998) such that
\begin{equation}
h \equiv \frac{1}{z}\times\frac{f}{g}.
\end{equation}
With respect to a particular reference cosmology, this becomes
\begin{equation}
h = \frac{f/g}{f_{0}/g_{0}}.
\label{eqn:h}
\end{equation}
To compute
$f_{0}$ and $g_{0}$ we use as a reference cosmology 
the same cosmology used in the Hubble volume 
simulation ($\Omega_{m} = 0.3, \Omega_{\Lambda}=0.7$).
 Throughout the paper we use this $h$ as the distortion
parameter unless noted
otherwise. Note that $\frac{\sigma}{\pi}$/$\frac{\sigma_{0}}{\pi_{0}} =
\frac{f}{g}$/$\frac{f_{0}}{g_{0}}$ because $\Delta \theta$ and $\Delta z$ 
represent the observed quantities and are invariant with cosmology. Therefore, 
$h$ is defined in such a way that when $h$ is greater (less) than 1, 
the correlation function is stretched in the transverse 
(line of sight) direction. 

For a given power-law correlation function used for $r \leq r_{0}$,
we adjust the amplitude of the CDM
correlation function  
so that it joins the power law exactly at  $r=r_{0}$. 
An alternative would be to keep the amplitude of the CDM
correlation function fixed but rescale $r$ so that it 
joins the power law. This is equivalent to assuming
a different value of $\Omega h$. We briefly describe
an analysis in \S3 where this was done in order to test whether a 
better fit to the non-linear correlation function could be achieved. In our
fiducial analysis, however we restrict ourselves to only changing
the amplitude of the CDM $\xi$. We note that as 
$r_{0}$ determines the change in the amplitude, using the CDM
shape does not increase the
number of free parameters in the fit. 

In redshift space, we observe a correlation function 
that is distorted because of
peculiar velocities and other noise sources such as 
redshift errors. We assume that the
distortion due to peculiar velocities can be removed
 using a measured  kSZ velocity (we return to the validity of this assumption 
in \S3.1) and we use a
Gaussian function of width $\sigma_{v}$ to model the
 remaining distortion.
The distortions only occur in the line of sight direction so
the effect can incorporated into the correlation function using a convolution
in $\pi$. Using the notation 
of Dalton \etal (1992),
the distorted correlation function for $r \leq r_{0}$,
can be written as: 
\begin{equation}
\xi(\sigma, \pi)_{r \leq r_{0}} = \frac{r_{0}^{-\gamma}}{\sqrt{{2\pi}}\sigma_{v}} \int
  \left[\sigma^{2}+\left(\pi-\frac{w}{H} \right)^{2} \right]^{-\gamma}
  e^{-\frac{w^{2}}{2 \sigma_{v}^{2}}} dw
\label{eqn:dalton1}
\end{equation}
When $r>r_{0}$, the CDM part of correlation function is convolved in the same
way:
\begin{equation}
\xi(r)_{r>r_{0}} = \frac{1}{\sqrt{{2\pi}}\sigma_{v}} \int \xi_{\rm CDM}(r)
  e^{-\frac{w^{2}}{2 \sigma_{v}^{2}}} dw
\label{eqn:dalton2}
\end{equation}
where $r = (\sigma^{2} + (\pi-\frac{w}{H} )^{2})^{1/2}$.

Our aim is to 
vary $h$, $\gamma$, $r_{0}$ and $\sigma_{v}$ and compare our model
 correlation function with that measured from a simulation.
This is to see if we can 
measure the distortion parameter $h$. Since $h$ is calculated with respect to
the $\Omega_{m} = 0.3, \Omega_{\Lambda} = 0.7$ cosmology (Equation 8),
we hope to recover $h=1$, thus  showing 
 that we can recover the cosmology that was used to generate 
the $\Lambda$CDM Hubble volume simulation. 
\subsection{Simulations}

We use a light-cone output of the $\Lambda$CDM 
Hubble Volume $N$-body simulation made publically
available by the Virgo 
Consortium (Frenk \etal 2000). The output is of a
model with $\Omega_{m} = 0.3, \Omega_{\Lambda} = 0.7, h = 0.7$ and
amplitude of mass fluctuations $\sigma_{8} = 0.9$.
The light cone data is an octant of the simulation volume 
in shape (and therefore covers $\pi/2$ steradian), with radius 3000 $\hmpc$.
 The redshift extends to $z=1.46$. We use the Virgo consortium
 cluster catalogue 
generated by using the spherical overdensity (SO) 
algorithm (Lacey \& Cole 1994)
with density threshold of 200, and a minimum particle count per cluster of 12
(Evrard \etal 2002).  The mass of each particle is 2.2$\times10^{12}\hmsun$ and
there are total of 802461 clusters with mass $> 2.6 \times 10^{13} \hmsun$
in the catalogue. We take this data set to be our fiducial ``simulated survey''
i.e. a survey that covers a quarter of the sky, down to this mass limit.

We calculate the cluster correlation function using the following estimator:
\begin{equation}
\xi(r) = C\frac{N_{cc}}{N_{cr}} -1
\label{eqn:xis_r}
\end{equation}
where $N_{cc}$ is the number of cluster-cluster pairs and $N_{cr}$ is the
number of cluster-random pairs. $C$ is the ratio of the number of clusters in
a random catalogue to the number of clusters in the cluster catalogue. Our
random catalogue has $C=20$. 
The random catalogues are made to have the same overall
number density with redshift trends as the simulation data and the same
survey boundaries. In order to do this, we bin the simulated clusters in
 redshift and use this to generate
a number density profile as a function of
redshift. 
We then randomly sample from this number density profile until 
we reach the correct total number of random points.

In order to measure the geometric distortion as a function of redshift,
we divide the simulated universe into 11 redshift bins from $z=0.2$
 to $z=1.3$. 
The space density of clusters declines at higher redshifts.
For example, we have space densities of  $3.5\times10^{-5},
6.0\times10^{-6}, 3.9\times 10^{-6}$ clusters per comoving unit volume ($\invhmpccc$)
for $z=0.2-0.3, 0.7-0.8,$ and $1.2-1.3$
respectively (24723, 83441 and 58767 clusters in the bins.)

 For each redshift bin, we compute $\xi(\sigma, \pi)$, in  2 $\hmpc$
intervals. In our fitting, we keep the 
494 bins in $\xi(\sigma, \pi)$ which have $r<50\hmpc$. 
For $\chi^{2}$ fitting,  we build a covariance matrix  using the
jack-knife method with 9 subvolumes of equal size. We only use the diagonal
elements in the covariance matrix for our analysis. This approach is
intermediate between using Poisson errors and a full covariance matrix.
We do not do the latter because with a small number of subsamples the
matrix becomes singular. Because galaxy clusters form a sparse sample,
this simplification of the error bar calculation is reasonable (see e.g.
Popowski and Weinberg 1998). We  check on this by computing the  $\chi^{2}$
per bin and comparing it to unity.

\subsection{Distortion in Redshift Space}

Observing clusters in redshift space means that cluster peculiar
motions cause distortion in the correlation function. 
Matsubara \& Suto (1996) modeled the redshift distortion of parallel and
transverse components of the correlation function, and
Padilla \& Baugh (2002) studied the redshift distortions
in the cluster correlation function using the Hubble volume simulation.

Peculiar velocities reflect large-scale cluster infall into 
overdense regions and result in the
correlation function being squashed in the line 
of sight direction. In the linear regime, the flattening effect is typically
described in terms of $\beta = \Omega^{0.6}_{m}/b$ where $b$ is the bias 
parameter.
 The geometric distortion and the 
peculiar velocity distortion both have roughly similar aspects,
although they can be distinguished given large enough
datasets or through their differing evolution with
redshift (see e.g., Ballinger \etal 1996).
 
\begin{figure}
\begin{center}
\epsfxsize=8.5cm
\epsfbox{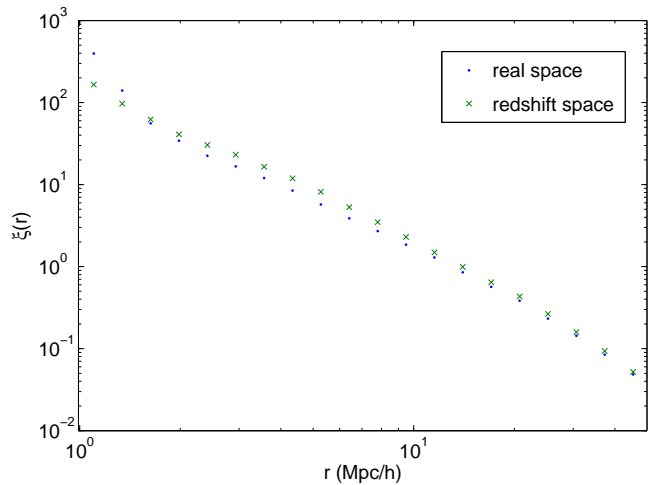}
\caption{
The simulation correlation function $\xi(r)$ 
for the Hubble volume octant in real and redshift space
for the redshift range $z=0.8-0.9.$ The difference between
the two in the linear regime agrees well
with the expected Kaiser (1987) enhancement in
redshift space (see text.)}
\label{fig:nu_rco}
\end{center}
\end{figure}

In Figure~\ref{fig:nu_rco} we plot the correlation function as a function
of $r$ in both real ($\xi_{r}$) and redshift space ($\xi_{s}$).
We see the suppression of correlation function on
small scales and the boost on linear scales as predicted
by Kaiser (1987). The linear theory prediction is
$\xi_{s}/\xi_{r}=1+\frac{2/3}\beta+\frac{1/5}\beta^{2}$, 
where $\beta=\Omega^{0.6}_m /b$,
with $b$ being the bias factor for clusters. Colberg
  \etal (2000) reported that $b$ for these
clusters is $b=2.25$, so that the linear boost correction factor due to 
Kaiser effect for $\Lambda$CDM is 1.150.  We average over the points in 
the linear regime with $r> 10 \hmpc$ and find an average boost of 1.146, close
to the linear theory prediction. 

Although we assume that the peculiar velocites can be removed via kSZ surveys,
we still have to deal with systematic or measurement errors that 
can not be eliminated entirely. Unlike peculiar velocities, measurement errors
will tend 
to stretch the correlation function along the line of sight. Padilla and Baugh
(2002) showed that the cluster correlation function is not as
subject as the galaxy correlation function to small scale 
virialized motions  (``the finger of God'' effect),
 due to the fact that superclusters are not virialized systems.

\begin{figure}
\begin{center}
\epsfxsize=8.5cm
\epsfbox{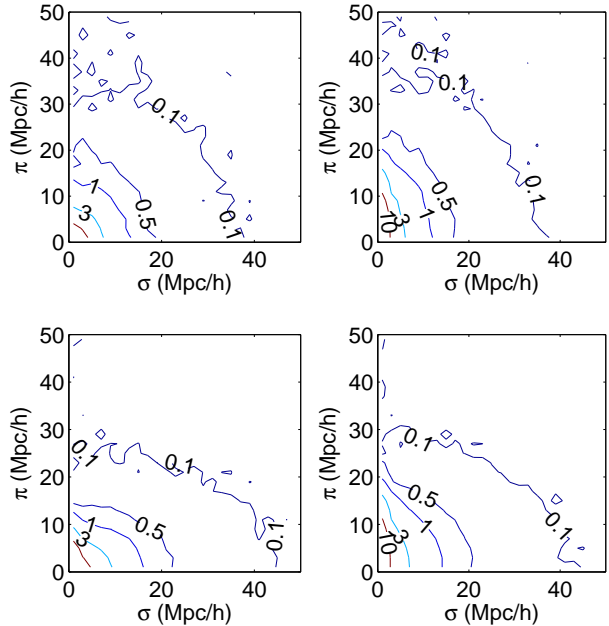}
\caption{Measured correlation functions
from the Hubble Volume simulation
for the redshift range $z=0.8-0.9$. Left (right) panels: without (with)
  Gaussian 
  measurement errors of 400 $km/s$. Top (bottom) panels: without (with)
  peculiar velocities.  The correlation function is almost isotropic when
  there is no 
  redshift distortion (top left). }
\label{fig:comp_ps}
\end{center}
\end{figure}

In Figure~\ref{fig:comp_ps} we examine $\xi(\sigma,\pi)$ measured
from the simulation.
We show how the correlation function is
distorted in redshift space when there are random errors (two right
panels) and when there are peculiar velocities (two bottom panels). For 
the random errors, we add Gaussian redshift 
errors with $\sigma=400$km/s
randomly to each cluster. The top left panel
shows the case when
there is no redshift distortion at all. The correlation function is 
isotropic, as expected. With random errors only, it
is stretched along the line of sight (top right), and undergoes further
distortion when peculiar velocities are added
(bottom right). The overall squashing along
the line of sight is better seen in the bottom left panel where peculiar
velocities are only source of distortion (no random errors added).
This agrees well with the results of Padilla \& Baugh (2002) who find  
that there is no stretching of the correlation function due to 
random peculiar
velocities in redshift space, but that 
coherent motions of clusters flatten the 
correlation function along the line of sight. 

Nagai \etal (2003) studied the real and estimated peculiar
velocities and their discrepancies with a hydrodynamical simulation. 
We describe their work in more detail in \S2.5 below. We take
their results to be representative
of the minimum noise level which can be 
achieved in  measurements of the cluster velocity. 

\subsection{SZ effect}

The SZ effect comprises the thermal and kinetic effects.
The thermal SZ effect is a distortion of the CMB photon
energy spectrum  due to
  inverse Compton
scattering in the cluster. When the CMB photons are scattered by free
electrons in the hot intracluster gas, they exchange energy and the tempature
decreases (increases) 
at low (high) frequencies, resulting in
a spectral distortion skewed toward higher frequencies.
 The distortion due to the thermal effect (non-relativistic) is described by:
\begin{equation}
\frac{\Delta T_{\rm tSZ}}{T_{CMB}}=
 y\left(x \frac{e^{x}+1}{e^{x}-1}-4 \right)
\end{equation}
where $x \equiv \frac {h \nu}{k_{B} T_{CMB}}$ and
\begin{equation}
y = \frac{k\sigma_{T}}{m_{e}c^{2}} \int dl T_{e}(l) n_{e} \approx \tau_{e}
 \frac{k
  T_{e}}{m_{e} c^{2}}.
\end{equation}

In addition, when the cluster is moving with respect to the CMB rest frame,
the signal will be Doppler shifted, with an amplitude  proportional to
the peculiar velocity of the cluster. This is the kinetic SZ effect.
The magnitude of the effect is given by
\begin{equation}
\frac{\Delta T_{\rm kSZ}}{T_{CMB}}= - \tau_{} \left( \frac{v_{pec}}{c} \right)
\end{equation}
where $\tau_{e}$ is the optical depth and $v_{pec}$ is the peculiar velocity
component along the line of sight. The clusters moving toward (away from) 
the observer have a negative (positive) sign for the
temperature decrement. For detailed reviews of the SZ effects,
see e.g., Sunyaev \& Zel'dovich (1980), Rephaeli (1995), Birkinshaw (1999),
Carlstrom \etal (2002). 

As Holder (2004) emphasized, only 1 \% of CMB photons interact with
free electrons in the hot cluster gas and the energy exchange per scattering
is another 1 \%, which makes the total thermal effect on
the order of $10^{-4}$. The kinetic effect is one magnitude smaller than
this ($10^{-5}$). Although both effects are very small, they have
different spectral signatures. For the thermal SZ, the intensity
decreases (increases) at low (high) frequencies, vanishing at
$\sim$ 220 GHz, whereas at the same frequency the kinetic effect is at
its peak and it has constant sign (positive/negative) throughout all
frequencies. With a multiwavelength SZ survey, we expect to 
be able to measure both the thermal and
kinetic effects, from which we can determine three physical properties of
each cluster ($\tau$, $v_{pec}$ and $T_{e}$). 

The cluster temperature can be obtained from X-ray observations
 or   from thermal SZ measurements
when the relativistic corrections are included (Pointecouteau
\etal 1998, Hansen \etal 2002). Sehgal \etal (2005)
 pointed out however that the SZ
measurements are not sufficient to 
break the degeneracy between the 3 cluster parameters ($\tau$, $v_{pec}$ and 
$T_{e}$) and
showed that an independent measurement of cluster temperature 
would greatly help.
 However, X-ray observations may not guarantee accurate information
on the cluster temperature either. One difficulty stems from
 the fact the X-ray
emission signal decreases with distance due to 
the inverse square law, making measurements at higher $z$ much harder
  (unlike the SZ effects). An additional problem arises because in
inhomogeous cluster gas, the 
temperature inferred from X-ray emission
 is affected by the clumpiness of the intra cluster gas while
it is irrelevant in thermal SZ measurements (e.g., Hansen 2004).

A recent study of systematic effects was carried out by by Diaferio etal 
(2005), who used hydrodynamic cosmological simulations to
create large samples of simulated clusters. It was found that it is
crucial to use the electron weighted cluster temperature, $T_{e}$ to recover
the peculiar velocity from the kSZ effect. Using the X-ray 
emission-weighted temperature, $T_{X}$
 can overestimate the peculiar velocity by 
$20-50\%$. Spatially resolved nearby clusters can be used to measure
$T_{e}$ in the center where it is comparable to $T_{X}$. However,
spatial modelling of the X-ray emission is still needed to seperate
the SZ effects, resulting in potential overestimate of the peculiar velocity
by $10-20 \%$.

The accuracy of $v_{pec}$ measurements depends on how well one can
reduce the systematic errors. Many studies (e.g. Knox \etal 2004) 
have been carried out testing how well one can hope to 
extract $v_{pec}$ via the kSZ effect, including the
effects of known systematic errors and noise
sources, e.g., interstellar dust emission, 
infra-red (IR) galaxies, radio sources,
imperfect bulk velocity, etc. The primary CMB
fluctuations which also impede accurate measurement of the
 kSZ because they have the
same spectral behavior (see Haehnelt \& Tegmark 1996, Aghanim \etal 2001,
Forgi \& Aghanim 2004, Aghanim \etal 2005). 
Some studies have estimated the potential 
measurement error on $v_{pec}$  to be
as low as $\sim100  kms$ (Nagai \etal 2004, Holder 2004, Sehgal \etal 2005),
while Benson \etal (2003) have set an observational 
upper limit on cluster velocities with errors of the order of $\sim1500
kms$. 
Our study is based on the former estimates, however we have also 
tried $\sim2000  kms$ as a measurement error. In this case we found
that the error bars were too large to be useful for constraining cosmological
parameters and we could not recover the distortion parameter ($h$) sensible
enough to continue for further study. If the measurement
errors are too big, we can not tell 
wheather the distortion is due to $\lambda$ or $v_{pec}$. This may be the most
common but challengin problem in AP test along with the fact that we need so many
accurate measurements of the clusters velocities.

\subsection{kSZ Errors from Hydrodynamic Simulations}

The measurement of peculiar  velocity from the kSZ effect
 in observational data will be affected by many 
sources of noise.
In our model fitting, we use a  Gaussian function to 
parametrize the statistical scatter on measurement
results. In creating the simulated surveys
from the Hubble volume simulation we will include the
option of using a more realistic noise distribution.
In this section, we use the error distributions
from the hydrodynamic simulation work of
 Nagai \etal (2003). 

These authors
use a high-resolution simulation of a galaxy 
cluster  by Kravtsov (2002)  to generate detailed kSZ
maps in a $\Lambda$CDM cosmology with $\Omega_{m} = 0.3,
\Omega_{\Lambda} = 0.7, h=0.7$. This is an 
$N$-body+gasdynamics simulations, but 
does not include the effects of magnetic fields, gas cooling, stellar feedback
or thermal conduction.  
Nagai \etal use their simulation to create simulated kSZ maps of the
cluster. They then analyse
these simulated maps in the same way one would with real
observational data, and use this to compute an kSZ-inferred cluster peculiar
velocity. This value is then compared to the known peculiar velocity of
the cluster in the simulation taken from the particle data. The idea
behind the paper is to see how well the two values match up, what the
scatter and systematic differences between the two are, and how the
measurement can be optimized,
For example, for the simulated kSZ maps, Nagai \etal try calculating
 the density-weighted velocities within certain
radii. They find that different aperture sizes do lead to different results
for the scatter between the true and inferred velocity.
This is because clusters are not perfectly spherical and have 
internal structures and motions.
If the aperture size is too small it may not reflect the bulk motion of entire 
cluster properly. We use the 
data calculated using the virial radius, which they find
is the most optimal case. 

A set of estimated and observed values of kSZ
velocities was kindly provided by Daisuke Nagai.
The simulated observations were generated for three
orthogonal projection angles and at nine different epochs ($a = 0.60, 0.65,
..., 0.95, 1.0$ where $a = 1/(1+z)$.) We will assume that these orthogonal
projections represent independent measurements.
 The peculiar velocity measurement errors are
the difference between the estimated and observed radial velocities. 
The distribution of $v^{est}_{r}-v^{obs}_{r}$ 
that we use is shown in Figure 6 of Nagai \etal. The standard deviation
of these measurement errors is $\sigma=50  \kms$.

For each of the 
clusters in our Hubble volume simulation catalogue,
we randomly pick a value from this set of adopted measurement errors, at
the closest appropriate redshift.
We then add it to the cluster's line of sight Hubble velocity and 
compute the new redshift.
We next convert the cluster coordinates into comoving coordinates assuming
an LCDM cosmology and compute the correlation function as normal.
As a result, we have a cluster correlation function distorted 
by measurement errors, one that would be seen from a survey
if the only source of error was due to differences between the
projected and actual cluster velocity. This represents the
best case that  could be obtained from a survey.  In \S4, however,
we will use larger and more realistic Gaussian errors when we simulate ACT or
SPT-like surveys. 

As mentioned above in reference to the work of Diaferio \etal 2005 and others,
systematic biases in measured velocities, such as over or underestimates
are also expected to occur in real measurements. Although in our fiducial 
modelling, we use the errors from bulk flow modelling, we will
also estimate the effects of systematic biases on cosmic geometry.

\section{Method}

In order to constrain cosmic geometry from the simulated cluster
catalogue, we fit our model correlation function (\S2.1) to it,
  varying four free parameters,  $h$,  $r_{0}$, $\gamma$, and $\sigma_{v}$.
 The parameter $r_{0}$ determines the overall amplitude of 
the correlation function,  and $\gamma$ its slope. The 
line of sight velocity distribution added to model measurement
errors is given by $\sigma_{v}$. The geometric
distortion, stretch in the transverse direction is quantified
by $h$. Our
purpose is to measure the distortion parameter $h$ 
and its uncertainty as a function of redshift by marginalizing over
the other three parameters. We then use $h(z)$ to 
 find the corresponding cosmology. 
The  detailed procedure is explained in the next section.

\subsection{Model fitting}
\begin{figure}
\begin{center}
\epsfxsize=8.5cm
\epsfbox{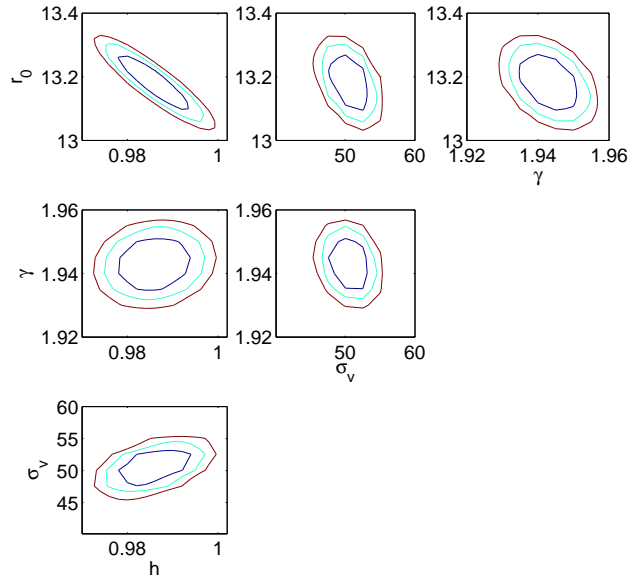}
\caption{Contours of $\Delta \chi^{2}$ 
for fits of the parameters $h,r_{0},\gamma$ and $\sigma_{v}$ for z=0.8-0.9.
 We fit the 2 point
correlation function 
from the Hubble volume simulation to the power law + CDM correlation function
(Eqn~\ref{eqn:dalton1}\&~\ref{eqn:dalton2}). The measurement errors from
the hydrodynamic simulation of Nagai \etal (2003) have been added. 
  $\Delta\chi^{2}$ in 4 parameter space has been marginalized for the two
  parameters of interest in each panel. Contour
  lines represent 1, 2 and 3 $\sigma$
confidence levels on  each of 2 parameters. }
\label{fig:6hyd}
\end{center}
\end{figure}

The model correlation function will be isotropic
when there is no redshift distortion due to peculiar velocities or measurement
errors. However, the measured correlation function will not be. The
two more parameters in addition to $r_{0}, \gamma$ which
we have introduced will parametrize the isotropy of
$\xi(\sigma, \pi)$. This is done by fitting the
measured correlation function with the distortion parameter $h$. In principle,
this should yield $h=1$. 
 However,
sampling variations will yield a measured 
correlation  that is not perfectly isotropic. The 
statistical uncertainty
 on $h$ measured from a particular catalogue will depend largely
on these sampling variations.

\begin{figure}
\begin{center}
\epsfxsize=9.0cm
\epsfbox{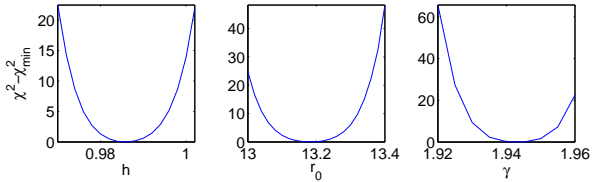}
\caption{$\Delta \chi^{2}$ for
fits of the geometric distortion parameter $h$, and 
the two correlation function parameters $r$ and $\gamma$
after marginalization over the other three
parameters in each case (see \S 3.1). 
}
\label{fig:3hyd}
\end{center}
\end{figure}

The fitting is done for $z = 0.2 \sim 1.4$ and $r = 1 \sim
50\hmpc$ and the procedure is as follows:
\begin{enumerate}
\item Add measurement errors to the cluster redshifts in the simulated 
catalog.
\item Convert angular positions and redshifts of clusters to comoving
coordinates assuming an LCDM cosmology.
\item Calculate the correlation function.
\item Compute a model correlation function parametrised 
by values of $h$, $r_{0}$, $\gamma$, $\sigma_{v}$
\item Calculate $\chi^{2}$ for the fit using the covariance matrix.
\item Marginalize the error in the parameter of interest.
\item Find the best fit parameters that yield 
  the minimum $\chi^{2}$.
\end{enumerate}
Since we calculate $\chi^{2}$ with 4 parameters, 
we need to marginalize it for each parameter. For example, to marginalize
$\chi^{2}$ in $h$ we rewrite $\chi^{2}$ in terms of the likelihood,
\begin{equation}
\chi^{2} = -2 \ln\mathcal{L}
\end{equation}
then 
\begin{equation}
\mathcal{L} = \exp^{-\frac{\chi^{2}}{2}}
\end{equation}
with this, calculate the average likelihood over $r_{0}$, $\gamma$ and
$\sigma_{v}$ 
\begin{equation}
\bar{\mathcal{L}} = \sum_{r_{0}, \gamma, \sigma_{v}} \exp^{-\frac{\chi^{2}}{2}} \Delta r_{0} \Delta \gamma \Delta \sigma_{v}
\label{eqn:3p}
\end{equation}
where $\Delta r$=$0.02  \hmpc$, $\Delta\gamma$=$0.005$ and $\Delta\sigma_{v}$=$2.5  \kms$.
Now we calculate  $\chi^{2}$ for $h$ from $\bar{\mathcal{L}}$
\begin{equation}
\chi^{2}_{h} = -2 \ln \bar{\mathcal{L}}.
\end{equation}
\begin{figure}
\begin{center}
\epsfxsize=8.5cm
\epsfbox{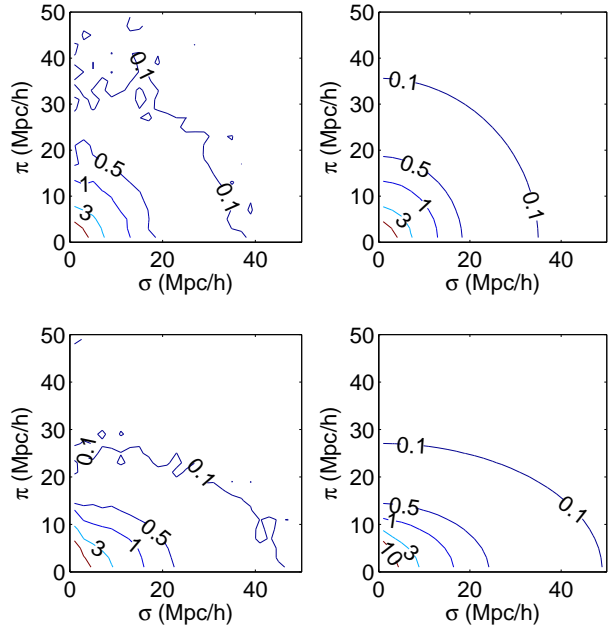}
\caption{Measured (left) and best fit (right) correlation
  functions for
the Hubble volume octant between redshifts $z=0.8$ and $z=0.9$ 
(this is $\sim 80000$ clusters).
 Measurement errors from the hydrodynamic simulation have been added.
  The effect of these  is not as visible as in Figure~\ref{fig:comp_ps} where
  much larger errors are added (400 $km/s$). For the bottom panels, peculiar
  velocities are also 
  included in addition to the measurement errors.}
\label{fig:4psbxi}
\end{center}
\end{figure}

In order to demonstrate our fitting procedure,
 we have applied it to one redshift bin in 
a simulated survey. This survey is the fiducial one, including 
all the clusters in the Hubble volume octant. The particular
redshift interval we choose is $z = 0.8 - 0.9$ and
includes $\sim 80000$ clusters.
In this example, we have applied the hydrodynamic simulation
velocity errors only to the cluster redshifts, so that we expect 
to recover a low value of the velocity dispersion parameter $\sigma$.

Figure~\ref{fig:6hyd} shows the resulting $\Delta \chi^{2}$ 
 ($\Delta \chi^{2} = \chi^{2} - \chi^{2}_{min}$) in two-parameter space,
with separate plots for each pair of parameters.
In each of the panels, the curves represent the $\Delta \chi^{2}$
values after marginalization over the other two parameters. 
The one dimensional distributions, from  $\Delta \chi^{2}$ values marginalized
over the 3 other parameters  (equation~\ref{eqn:3p})are shown in 
Fig.~\ref{fig:3hyd}. We
find 1$\sigma$ statistical errors on $h, r_{0}$ and $\gamma$ of
 0.5\%, 0.4\%, and 0.3\% respectively. The important parameter for
our purposes is $h$, for which we find a best fit value of $0.987$,
which is $1.3 \%$ ($\sim 2.5  \sigma$) from the expected $h=1$.  
We attribute this percent level bias to the fact that the model correlation
function we fit with is not a perfect match to the simulation correlation
function. This type of systematic bias is likely to be difficult
to circumvent. We return to this in the discussion (\S5.2).
The chi squared value for the best fit in this case is 578, for 494 bins,
so while the fit is good, there is room for slight improvement.

We examine the $\xi(\sigma,\pi)$ plots from the simulation in
Fig.~\ref{fig:4psbxi}, alongside the best fitting model $\xi$. 
In the top panel of Fig. ~\ref{fig:4psbxi}
we have included only the measurement errors
from hydrodynamic simulations, but assumed that the peculiar velocities
have been subtracted.  Although the velocity error
distribtion is not Gaussian, we have fitted it in our
convolution using a Gaussian function (equations~\ref{eqn:dalton1} \&~\ref{eqn:dalton2}),
which has a best fitting $\sigma\sim 50 km/s$ (see fig~\ref{fig:6hyd}).
By eye, it is difficult to see effect of these velocity errors
making the  $\xi(\sigma,\pi)$ anisotropic, although this effect
sucessfully recovered in the fitting.

In order to judge the effect of peculiar velocities on a straight fit
to the cosmic geometry, we have also tried fitting our 4 parameters
without subtracting the peculiar velocities. By doing this, we will get an
idea of how the peculiar velocities can mimic the effect of geometric
distortions. This was addressed in the context of linear theory
by Ballinger \etal (1996), who found the relationship
governing the effective value of $h$ is $h \sim 1+2\beta$.
Here $\beta=\Omega_{m}^{0.6}/b$, with $b$ being the bias parameter. In 
the present case, $b$  (the ratio of the cluster $\xi$ to 
the matter $\xi$) is $\sim 2.5$, so that $\beta\sim 0.2-0.4$ over
redshifts from $z=0-1$. Based on this, we expect $h$ to be $\sim 1.4-1.8$
when peculiar velocities are not removed. 

\begin{figure}
\begin{center}
\epsfxsize=8.5cm
\epsfbox{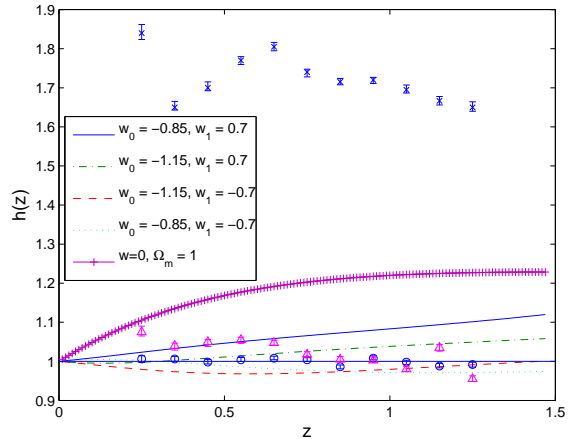}
\caption{Distortion parameter  
$h$ vs. $z$ for different cosmologies with repect to 
an $\Omega_{m} =0.3$, 
$\Omega_{\Lambda} = 0.7$ cosmology. We show theoretical results 
for a few example cosmologies as lines ($\Omega_{m}=0.3$ unless noted 
otherwise). For the simulations, we show results as circles  
(with 1 $\sigma$ error bars) for the  
case with peculiar velocities corrected when random errors from hydrodynamic
simulation are added. 
We also plot as triangles results for the simulations where 
much larger velocity errors
were added (a Gaussian $\sigma$ of 400$\kms$)
Finally, we also show results for 
the case where the peculiar velocities have been left in, as crosses,
which yields an extremely distorted correlation function ($h$ very different
from 1.) 
}
\label{fig:hze}
\end{center}
\end{figure}

This simulation test was carried out,
and  $\xi(\sigma,\pi)$ is shown in the bottom panels of Fig.
 ~\ref{fig:4psbxi}, alongside the best fitting model which includes
geometric distortions only. We can see that the peculiar velocities
cause a very strong distortion, and that the form of the
distortion is not obviously different to the eye when compared
to the geometric one, a point which has been made by many authors including
Ballinger \etal (1996) and Matsubara and Suto (1996).
When we fit the $h(z)$ that results
(Fig.~\ref{fig:hze}) we find values $\sim 1.6-1.8$, as expected.
The $\chi^{2}$ per bin is 3.2 (1616 for 494 bins), which is
 substantially worse than the
fit to true geometric distortions.
This difference in the fit could be useful if accurate estimates of the
peculiar velocities are not removed.

For example, a systematic overestimate of the velocity will cause
the distortion $h(z)$ to be underestimated.
In the worst case scenario detailed  by Diaferio \etal 2005, 
using a naive measurement based on 
clusters unresolved in X-rays would result in an overestimate of
the peculiar velocity of $20-50\%$. Subtracting these cluster velocities
to recover an estimate of the real space correlation function 
would yield $h(z)$ values $\sim 0.1-0.35$ too low.
This is at the level which would make an Einstein De Sitter model
look like a concordance $\Lambda$ model, and so this would
not result in an acceptable test of cosmic geometry. Even with 
a spatially resolved measurement of the temperature, from  Diaferio \etal 2005
we would expect a bias in $h(z)$ of $\sim 5-10\%$, which would affect
a measurement of $\Omega_{\Lambda}$ at the $\sim 0.1-0.2$ level.
We note that Seghal \etal (2005) find somewhat more optimistic conclusions from
a simulation study of constraining cluster parameters from SZ
observations, finding that velocity errors could be biased by as
 low as $15-40 \kms$ if X-ray weighted temperature measurements are used.
This said, there are also some improvements which could be made
to reduce the bias, such as using the integrated SZ flux together with 
scaling relations (Benson \etal 2004), and/or
 making use of the $z$ dependence
of the distortion more efficiently. As we have seen above, it is also
possible that the relatively poor goodness of fit when 
peculiar velocity distortions remain will help diagnose this without
a theoretical model for the peculiar velocities being necessary.

 With presently planned surveys such
as ACT and SPT, these biases will be comparable to or smaller than the
statistical errors, however they will represent an
severe obstacle to any attempt to use the cluster real space AP test as 
a precise probe of cosmology.

In Fig.~\ref{fig:hze}
we also show $h$ vs. $z$ from our fitting to the simulation
after subtracting peculiar velocities, but with the
errors from hydrodynamic simulations.
We can see that the expected result for LCDM, $h=1$ is recovered well, with
error bars of $\sim 1 \%$. The trend of $h(z)$ expected  in some other 
example cosmologies is also shown as lines. The differences in $h(z)$ 
between an $\Omega_{m}=1$ model and LCDM are of the order of
10s of percent. Models with slightly different dark energy 
equations of state from LCDM fall even closer. We explore these below.

We have also carried out simulation tests using much larger 
peculiar velocity errors. For example, we have assigned random
Gaussian distributed
errors of $400 \kms$ to the clusters before carrying out our
fitting procedure (Figure 3). In this case, we find values of $h(z)$ which 
are approximately $3\%$ too high, when averaged over the 10 redshift bins.
This corresponds to a quite large bias, and means that it seems
to be difficult to recover the correct cosmic geometry with such 
large velocity errors. With errors $\sim 200 \kms$, we find that the
fitting technique works much better. Errors of this magnitude are
to be expected given good observational data (see e.g., Diaferio \etal 2005,
Seghal \etal 2005). We have not so far mentioned the errors on the 
cluster redshifts, but of course an additional complication is that they
must also be measured to a comparably high level of precision.

\subsection{Varying the dark energy equation of state}

We adopt a simple equation of state, $w = w_{0} + w_{1}z$,
where the pressure, $p=w\rho$. In order to find
the best fit values for $w_{0}$ and $w_{1}$, we vary $w_{0}$, $w_{1}$ and
$\Omega_{M}$ on a grid, and compute the expected $h(z)$ for each set 
of parameters. We then calculate $\Delta \chi^{2}$ by fitting to the
simulation results in each redshift bin. $\chi^{2}$ is given
by: 
\begin{equation}
\chi^{2} = \sum^{n}_{i=1} \frac{(h_{i} - h'_{i})^{2}}{\sigma^{2}}
\label{eqn:nchi}
\end{equation}
where $h_{i}$ represents the best-fit distortion parameter $h$ in redshift bin
$i$, $h'_{i}$ is the distortion parameter with given $w_{0}$, $w_{1}$
and $\Omega_{M}$, and the sum is over $n$ bins in redshift.
In Eqn~\ref{eqn:nchi}, the $\sigma$'s have been obtained from
the 1D marginalized results for $h(z)$ from the simulations (e.g. 
Figure 5). In the few cases where in the higher dimensional
space before marginalization, the  1$\sigma$ contour for $h$ is out of
the range, we are conservative and set the $\chi^{2}$ to a very large number so
that during the marginalization the contribution to the likelihood
from that point is negligible.

\begin{figure}
\begin{center}
\epsfxsize=6.50cm
\epsfbox{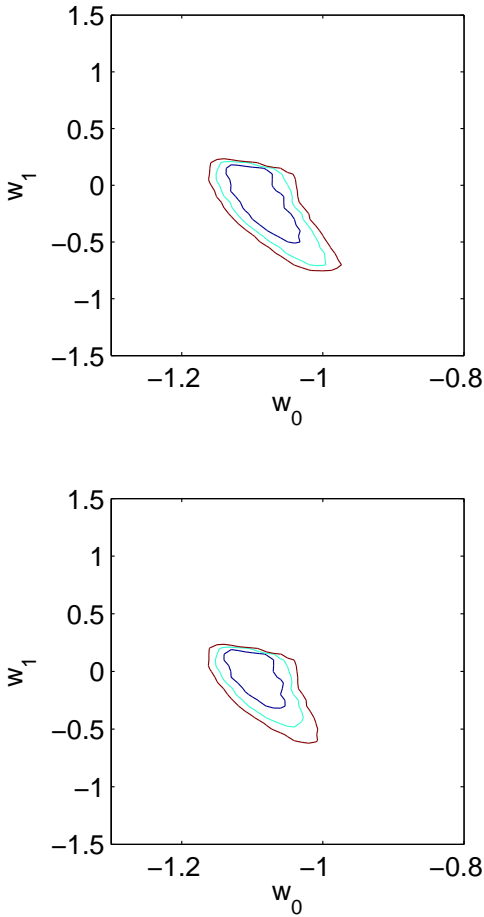}
\caption{Fitting cosmological 
equation of state 
parameters (see \S3.2) to the fiducial
Hubble Volume simulation catalogue. Best fit at $w_{0} = -1.08, w_{1} =
  -0.15$.  We show $1, 2$ and $3 \sigma$ contours. 
The top panel is without a prior on $\Omega_{m}$ and has a minimum at
  $\Omega_{m} = 0.32$. The bottom is with a prior of $\Omega_{m} = 0.3$,
with uncertainty on $\Omega_{m}$ of  $\sigma=0.05$.
For visual purposes, a  
Gaussian filter with $\sigma=0.8$ grid cells was used to 
smooth before plotting.
}
\label{fig:w0w1}
\end{center}
\end{figure}

Once we have a 3d grid of $\Delta \chi^{2}$ values for $w_{0}, w_{1}$ and
$\Omega_{M}$ we marginalize over $\Omega_{m}$. We do this either with
no external prior, or else after imposing  
a Gaussian prior on $\Omega_{m}$ ($\mathcal{L}' =
\mathcal{L} \exp^{-\frac{(\Omega_{m}-0.3)^2}{0.05}}$). We expect
to recover $w_{0} = -1$, $w_{1} = 0$, as this is
 the cosmology used in the Hubble volume
simulation.

\begin{figure}
\begin{center}
\epsfxsize=8.0cm
\epsfbox{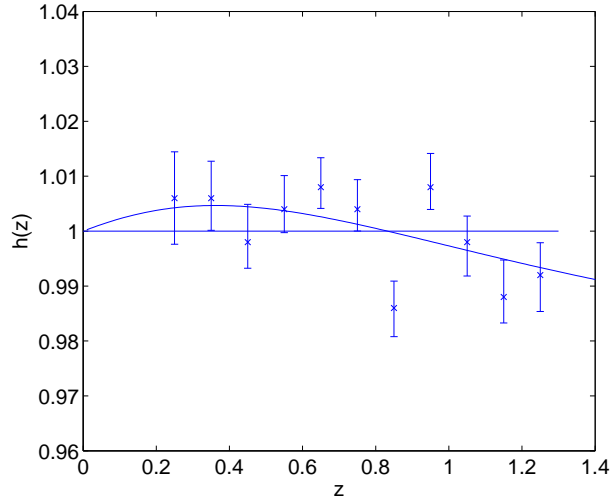}
\caption{Distortion parameter $h(z)$
measured from the fiducial Hubble Volume
 simulations (points) 
together with the theoretical curve for the 
  best fit $w_{0}, 
w_{1}$ and $\Omega_{m}$ (which gives the minimum $\chi^{2}$ in Figure 8.)
The line for the correct model $\Lambda$CDM
is $h(z)=1$.
}
\label{fig:hz_lcdm}
\end{center}
\end{figure}

From our fitting procedure we find 
$w_{0} = -1.09^{0.022}_{0.022}$ and $w_{1} = -0.15^{0.29}_{0.17}$
with no prior imposed on $\Omega_{m}$ ($1 \sigma$
error bars). 
Adding the prior makes little difference,
with the best fit being
 $w_{0} = -1.09^{0.015}_{0.026}$ and $w_{1} = 0.1^{0.05}_{0.27}$.
In Figure 8 we show the $\Delta \chi^{2}$ contours for these
two cases. 
 Although we have seen from 
Figure~\ref{fig:w0w1} that the distortion
parameter $h$ can be recovered at the  percent level, this still
translates into relatively loose constraints on $w_{1}$, 
regardless of the prior on $\Omega_{m}$.

In Figure \ref{fig:hz_lcdm} we show the shape of the
theoretical $h(z)$ curve for this best fitting cosmology, compared
to the measured values from the simulation. We can see that the
best fit curve follows the dip in $h(z)$ values at high z, which account
for the difference from the true cosmology.

\begin{figure}
\begin{center}
\epsfxsize=7.5cm
\epsfbox{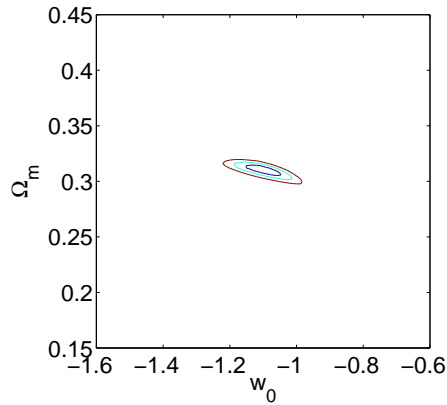}
\caption{Constraints on $w_{0}$ and $\Omega_{m}$ for the 
fiducial Hubble Volume simulation catalogue
case when we set
  $w_{1}=0$. We show $1, 2$ and $3 \sigma$ contours. 
The cases with and without a prior on
  $\Omega_{m}$ give almost the same result. Best fit at $w_{0}=-1.09$ and
 $\Omega_{m}= 0.31$.}
\label{fig:w0om}
\end{center}
\end{figure}

In Figure~\ref{fig:w0om} we set $w_{1}=0$,
in order to assess the type of constraints possible
in this case.  We
find a minimum $\chi^{2}$ 
at $\Omega_{m}= 0.31^{0.002}_{0.002}$, and
$w_{0}=-1.09^{0.020}_{0.037}$. Imposing the previous prior
on $\Omega_{m}$ makes no difference to this result.

\subsection{Test on the $\tau CDM$ simulation}

We now test our method on a different simulated
universe. In the previous case, have taken data
sets from $\Lambda$CDM simulations, and then
assumed the $\Lambda$CDM geometry when analyzing them.
Because of this, we expected to recover a distortion 
parameter $h(z)$=1. Now instead, we use a different
simulation, but still assume the $\Lambda$CDM geometry 
to analyze it. In this case, we expect to find $h(z)\neq 1$.

 For the simulated universe,
we use the Hubble volume simulation of the $\tau$CDM model
(see Frenk et al 2000). This
is a cosmology with $\Omega_{m} = 1,
\Omega_{\Lambda} = 0$. The linear  matter power spectrum has a 
similar shape to that of LCDM. We use all the clusters in an octant
as before, except this time the maximum cluster redshift is  
 $z_{max} = 1.25$.

\begin{figure}
\begin{center}
\epsfxsize=9.0cm
\epsfbox{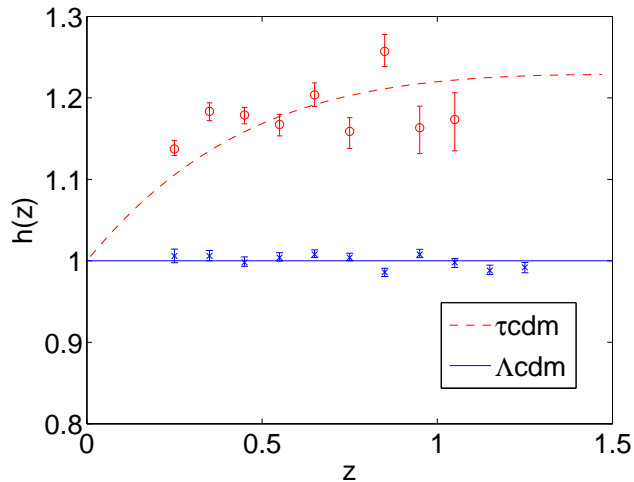}
\caption{
Theoretical
distortion parameter $h(z)$ 
 for $\tau$CDM and $\Lambda$CDM (lines). Data points with circles (crosses)
  represent fitted distortion parameters from the simulation 
with 1 $\sigma$ error bars for
  $\tau$CDM ($\Lambda$CDM).}
\label{fig:hz_all}
\end{center}
\end{figure}

We first compute the angular positions and redshifts
of the clusters, as they would be seen by an observer.
We then use the $\Lambda$CDM relations (Equations 5 and 6) to
convert these into comoving coordinates, assuming an observer at the origin.
We compute the correlation function from the simulation as before, 
and fit the distortion parameter $h(z)$.

This test is is analogous to the real situation,
 where we would have
observational data from an unknown cosmology and use an assumed cosmology
during the fitting procedure. The unknown cosmology here
 is an Einstein-de Sitter universe with $\Omega_{m}=1$, and we
find the distortion parameter with respect to a cosmology with $\Omega_{m} =
0.3, \Omega_{\Lambda} = 0.7$.

Using Equation 8 we compute the 
distortion parameter which we expect to recover in this
case. This is shown as a dotted line in Figure~\ref{fig:hz_all}. 
When we carry out the test, we find results which are intermediate 
between this line and the $h(z)=1$ LCDM line at all redshifts. This
is an indication that in a realistic situation such as our test, the
distortion parameter cannot be inferred simply when the 
assumed model (here LCDM) is very different from the actual Universe
(in this case $\tau$CDM). Instead, if the distortion parameter is found to be
 very different from $h=1$, an iterative approach should be used. In this
case, the assumed cosmology on the second try would be one intermediate 
between LCDM and  $\tau$CDM. By using this method, we would approach the
true cosmology, albeit more slowly.
In Figure~\ref{fig:hz_all} we show the recovery of the  Einstein-de Sitter
geometry as points with error bars, using  the EDS model as
an assumed cosmology. This is representative of what should
occur with the full iterative procedure, although this has not been fully 
tested in this case. 
 We have rescaled the result for $h(z)$ by the ratio
of the analytical predictions for $\Lambda$CDM and $\tau$CDM  so
that we can show the results on the same plot as the  $\Lambda$CDM
results. Although the fit is not quite as good as with the 
 $\Lambda$CDM  test, the results show that by moving towards an 
estimate of an assumed cosmology we can expect good results.

\begin{figure}
\begin{center}
\epsfxsize=8.0cm
\epsfbox{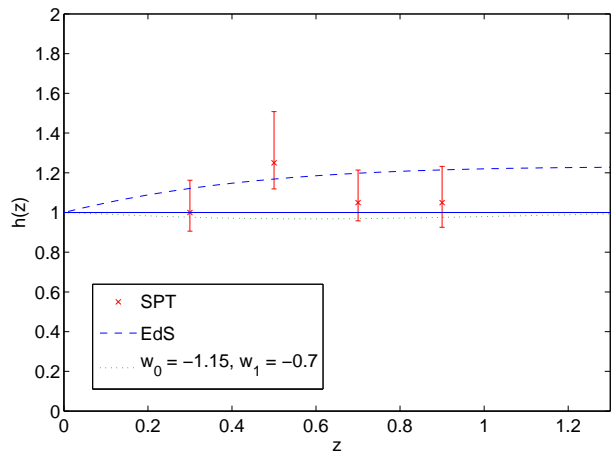}
\caption{Distortion parameter $h(z)$ with 1 $\sigma$ error bars
for the simulated SPT cluster survey. 
3 example theory
curves are shown as lines.
}
\label{fig:hz_spt}
\end{center}
\end{figure}

\section{Results from simulated observational catalogues}

\subsection{The South Pole Telescope}

The SPT survey
 will cover 4000 square degrees 
in 5 frequency bands 
and is 
expected to observe $\sim$20,000
clusters with masses greater than
2$\times 10^{14}\msun$. For our
SPT (and ACT, see below)
simulations, we use a Gaussian measurement error of 100 $km/s$, based
on published estimates of the best likely error on the kSZ velocity 
measurements (Nagai \etal 2004, Holder
2004, Sehgal \etal 2005). The standard deviation of errors from the 
hydrodynamic
simulation which we used before in the Hubble volume case was $\sim50  km/s$,
 so the present value gives slightly bigger velocity distortions.

The simulation output we use to make our mock catalogue is an
octant shape and covers almost the
same area (we again use the $\Lambda CDM$ simulation),
so to simulate the SPT survey we set a mass threshold so
that the number of clusters that we observe is around
20,000. This means that the total number of  clusters in the mock survey is
only 1/40 of that in the underlying simulation.
Since we have fewer clusters, we increase the size of each
redshift bin and include clusters up to $z=1$ only.
Our covariance matrix is constructed in 
the same way as in \S3.1. When building the covariance matrix, we make
sure that $\sigma, \pi$ bins are large enough  that there 
are cluster pairs in each bin, to avoid singular error bars.

 The results are shown in Figure ~\ref{fig:hz_spt}.
As expected, we have large error bars because of the
of the smaller number of clusters than in the fully sampled
simulation. The points also lie somewhat above the
$h(z)=1$ curve, but the results still give a reasonable
estimate of cosmic geometry without a large bias. 
Due to the large error bars, it is not possible
to constrain $w_{0}$-$w_{1}$ within reasonable error bounds.
However, by assuming that the dark energy density does not 
change with redshift, we can still find useful constraints on 
$\Omega_\Lambda$, and $w_{0}$, for example. In Figure \ref{fig:dchi_ol_act} we show
the constraints on $\Omega_\Lambda$ which result when a flat cosmology
is assumed and $w_{1}=0$. We find 1 sigma errors on $\Omega_\Lambda$
of $^{+0.14}_{-0.09}$, and a central value within 1 $\sigma$ of the
correct one. The constraints on $w_{0}$ (Figure \ref{fig:dchi_w0_act}) are weaker,
with $w_{0}<-0.76$ at 1 $\sigma$.

\begin{figure}
\begin{center}
\epsfxsize=8.0cm
\epsfbox{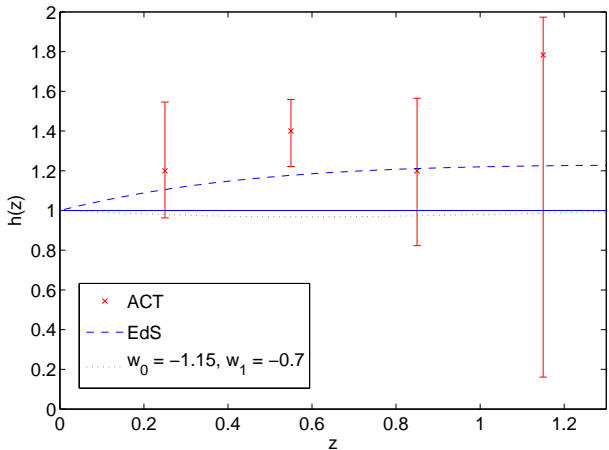}
\caption{Distortion parameter $h(z)$ with 1 $\sigma$ error bars
for the simulated ACT cluster survey. 
3 example theory
curves are shown as lines.
}
\label{fig:hz_act2}
\end{center}
\end{figure}

\subsection{The Atacama Cosmology Telescope}

The ACT
 will cover only 200 square degree of the sky, but
at higher sensitivity than the SPT. It is expected to observe 
$\sim 1000$
clusters with masses greater than $2\times10^{14}\msun$. 
With this mass threshold we find fewer cluster in 
the Hubble Volume simulation in a comparable area, so we use
$10^{14}\msun$ as our threshold. The $\chi^{2}$ analysis has been done in 
the same way as before with a Gaussian measurement error of 100 $kms$, except that the covariance matrix has been built
 using 6 separate volumes of 200 square
degrees taken from the Hubble Volume octant. 
We then take one subvolume as our simulated survey
and carry out our $\chi^{2}$ 
analysis using that.

\begin{figure}
\begin{center}
\epsfxsize=8.0cm
\epsfbox{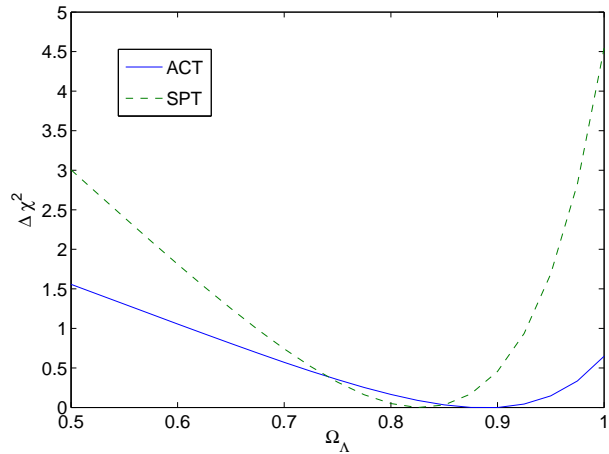}
\caption{
Constraints on $\Omega_{\Lambda}$ for 
mock ACT and mock SPT surveys, assuming that $w_{1}=0$ and $\Omega_{m} + 
\Omega_{\Lambda}= 1$. 
}
\label{fig:dchi_ol_act}
\end{center}
\end{figure}

Since we now have even fewer clusters
than in the mock SPT survey,
 the error bars are substantially larger (Figure ~\ref{fig:hz_act2}). It is
not possible to put any useful constraints in
the $w_{0}$-$w_{1}$ plane without other priors. 
Again we we try to see if it is feasible
to constrain $\Omega_{\Lambda}$ setting $w_{1} = 0$ (Figure
~\ref{fig:dchi_ol_act}) and $w_{0}$ (Figure 15). We find 
1 $\sigma$ errors on $\Omega_{\Lambda}$ $\sim 2$ times as
large as for the mock SPT survey and a $1 \sigma$ limit on $w_{0} < -0.45$.

From the analysis of ACT mock surveys we can see that the relatively small
cluster sample and the need to split it up into redshift bins
will make it difficult to measure the correlation
function accurately enough to make this type of measurement. Nevertheless,
some constraints will be possible, and the error bars of e.g. $\sim 0.2-0.3$
on $\Omega_{\Lambda}$ are still useful.

\begin{figure}
\begin{center}
\epsfxsize=8.0cm
\epsfbox{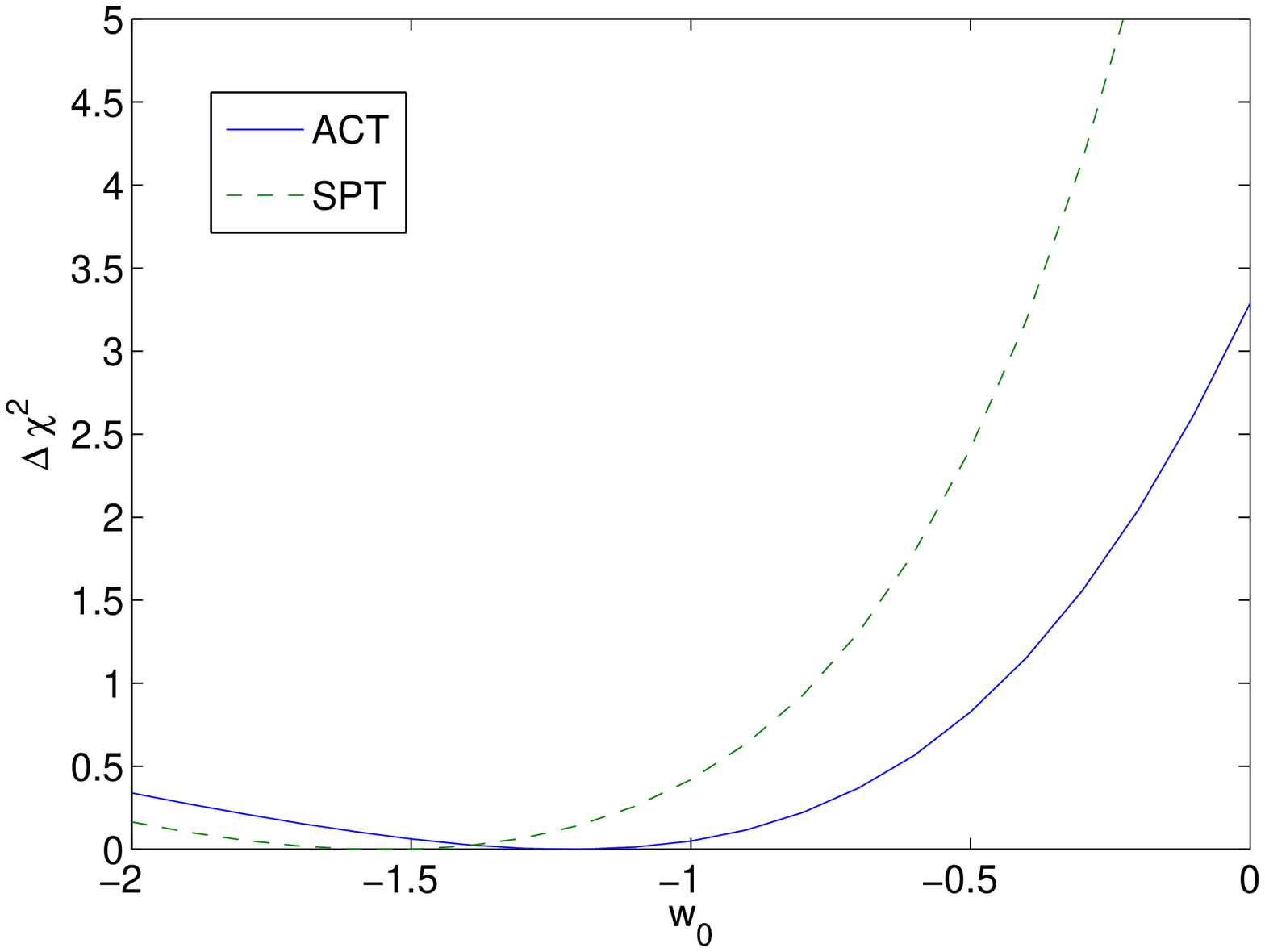}
\caption{
Constraints on $w_{0}$ for 
mock ACT and mock SPT surveys, assuming that $w_{1}=0$ and $\Omega_{m} + 
\Omega_{X}= 1$. 
}
\label{fig:dchi_w0_act}
\end{center}
\end{figure}

\section{Summary and Discussion}

\subsection{Summary}
Using a combination of the Hubble Volume simulation (Frenk \etal 1999) and
the hydrodynamic simulation results of Nagai \etal (2003), we have tested
using the peculiar velocity corrected
galaxy cluster correlation function to measure cosmic geometry. Our
findings can be summarised as follows:\\
\noindent(1) The galaxy cluster correlation function, corrected
using kSZ velocities can act as an intrinsically isotropic
object for the AP test.\\
(2) The geometric distortion of the correlation function needs to be 
measured to $\%$ level accuracy in several redshift bins to constrain the
equation of state parameter $w_{0}$ to better than $10 \%$.\\
(3) Our simulation tests assuming errors from hydrodynamic
modelling show that this can be done with a survey covering 1/8th of
the sky, with clusters out to $z\sim 1$.\\ 
(4) The distortion from peculiar velocities will be very difficult to 
correct. Systematic over or underestimates of the peculiar
velocities from kSZ measurements will be more problematic
than statistical errors.\\
(5) Of surveys currently underway or planned, the SPT seems to
be the most promising, and could yield constraints on $\Omega_{\Lambda}$ 
accurate to $10-20\%$.

\subsection{Discussion}

Of the many recent papers which have been written on using the AP
test (e.g. Mc Donald \etal 2003),
 all take the approach of modelling the peculiar
velocity component of the correlation function distortion. The fact that
for galaxy clusters the peculiar velocity can be instead measured and
subtracted offers the potential of a useful complementary method. For
example, in order to model  peculiar velocities well, the relationship
between the clustering of the objects and the underlying mass field
needs to be known to high precision. This is not necessary in the case
where the velocities are subtracted. The latter method does however
bring with it several new difficulties, including the
necessarily sparse nature of galaxy cluster samples and the extremely
challenging problem of measuring the peculiar 
velocities of thousands of clusters to high accuracy, when to date this
has not yet been done for a single cluster.

One of the ingredients in our measurement of the distortion is our
model for the correlation function as a function of separation $r$.
 We have found that using a fitting function
which does not model the correlation function well can lead to 
biases in the measured distortion.
For example, before adopting our fiducial fitting function,
 a power law combined
with a CDM linear theory $\xi$, we tried a using single
power law as the fitting function. This resulted in 
a systematic 10\% bias in the value of $h(z)$.
We also tried using a single power law  to fit $\xi$ to regions
$r=1 \sim 40 \hmpc$ (following da \^{A}ngela \etal 2005).
In this case the fitted $h$ in each redshift
bin changes slightly but the overall fitting of cosmological
 parameters does not improve very much. Using a more sophisticated
fitting function than a power law therefore seems to be crucial.

The spectral
distortion in the SZ effects is determined by $\tau$, $v_{pec}$ and $T_{e}$.
These three parameters could be disentangled using multiwavelength
observations. Sehgal \etal (2005) have shown how this will be
possible  in light of  current survey
sensitivities and suggest an independent temperature measurement from X-ray
observations in order to break  degeneracies.
 However,  as mentioned in \S2.5, the temperature
obtained by X-ray spectroscopy for unresolved clusters is not  
the electron- weighted temperature which is required to separate three
parameters ($\tau$, $v_{pec}$, $T_{e}$). Mathiesen \& Evrard (2001) estimated
that the discrepancy can be as much as 1keV.  
However, Sehgal \etal found that this
 temperature discrepancy will lead to a velocity
bias of 15-40 km/s. They also claim that for an ACT-like survey, a
2 keV temperature error can still only lead to peculiar velocities 
with errors less than 100 km/s. 
We have found that the statistical measurement errors of this size
can be modelled well, but that any systematic
bias in the velocities will give biased results for
cosmic geometry. This issue will need to be explored in more detail
once large samples of clusters with kSZ measurements are available.
Indeed, if the bias is too large, then it is possible that by assuming a 
model for cosmic geometry the distortion can be used instead to
calibrate the peculiar velocities.

There many other planned surveys than ACT and SPT
 that are designed to be sensitive to SZ signals from clusters. 
For example, AMiBA\footnote{http://amiba.asiaa.sinica.edu.tw},
 AMI\footnote{http://www.mrao.cam.ac.uk/telescopes/ami/}
 ( $\sim 100$ clusters), APEX\footnote{http://bolo.berkeley.edu/apexsz}
($\sim 1000$ clusters) and
Planck\footnote{http://www.rssd.esa.int/Planck} ($\sim 10000$ clusters).
The Planck mission promises the largest
survey area (the whole sky),  compared to
 SPT for example which will observe 20,000 clusters over
4000 square degrees (about 1/8 of the sky). Planck will however have
much larger CMB temperature measurement uncertainties,
and so will only observe a similar number of clusters. 
They will therefore be distributed even more scarcely,
making correlation function measurements difficult. We have tried sampling
our simulation down to the space density of Planck clusters, but find that the
Poisson errors on $\xi$ make it unfeasible to extract useful constraints
on the geometric distortion.

In conclusion, we have shown that by subtracting kSZ measured
velocities, it is possible to carry out a test of cosmic geometry
with an intrinsically isotropic object, as envisaged in the 
original paper by Alock and Paczynski (1979). Using simulations
we have seen that given small enough 
systematic and statistical errors, the correct geometry can
be recovered. However, huge samples of galaxy clusters are required
with precisely measured velocities. For example, to 
measure $w_{0}$ with 3\% statistical errors we have seen that 
800,000 clusters are necessary.
As the kSZ signals are so small, observing the kSZ effects from
this many clusters is not possible with currently planned surveys.
However more advanced surveys are
 planned to detect more clusters in the future and the test we have
proposed here will offer an additional way to use such data to
constrain dark energy.

\section*{Acknowledgments}
We thank the referee, Michael Strauss for a number of suggestions which 
improved the paper and Daisuke Nagai for providing the hydrodynamic
simulation results. RACC acknowledges
support from the NASA LTSA program, contract NAG5-11634.


\begin{thebibliography}{99}

\bibitem[]{aghanim1}
Aghanim N., G\'{o}rski, K.M., Puget, J.-L. 2001, A\&A, 374, 1., Lagache G., 2005, A\&A, 439, 901

\bibitem[]{aghanim2}
Aghanim N., Hansen S.H., Lagache G., 2005, A\&A, 439, 901

\bibitem[]{atrio}
Atrio-Barandela F., Kashlinsky A., Mücket J. P., 2004, \apj, 601, L111

\bibitem[]{ap}
Alcock C., Paczy\'{n}ski B., 1979, Nat, 281, 358


\bibitem[]{ballinger}
Ballinger W.E., Peacock J.A., Heavens A.F., 1996, \mnras 282, 877

\bibitem[]{bbks}
Bardeen, J., Bond,J.R., Kaiser, N. \& Szalay,A.S. 1986, \apj, 304, 15

\bibitem[]{benson}
Benson, B. A., Church, S. E., Ade, P. A. R., Bock,
 J. J., Ganga, K. M.,
 Hinderks, J. R., Mauskopf, P. D., Philhour, B., Runyan, M. C. 
\& Thompson, K. L. 2003, \apj, 592, 674

\bibitem[]{benson}
Benson, B. A., Church, S. E., Ade, P. A. R., Bock,
 J. J., Ganga, K. M., Henson, C. N. \& Thompson, K. L., 2004,
\apj, 617, 829

\bibitem[]{ballinger}
Birkinshaw, M., 1999, Phys. Rep., 310, 97

\bibitem[]{blain}
Blain, A.W., 1998, \mnras, 297, 502

\bibitem[]{borgani}
Borgani S. \etal, 1997, new Astron., 1, 321

\bibitem[]{carlstrom}
Carlstrom J.E., Holder G.P., Reese E.D., 2002, ARA\&A, 40, 643

\bibitem[]{colberg}
Colberg J. M., White S. D. M., Yoshida N., MacFarland T. J., Jenkins A.,
Frenk C. S., Pearce F. R., Evrard A. E., Couchman H. M. P., Efstathiou
G., Peacock J. A., Thomas P. A., 2000, \mnras, 319, 209 

\bibitem[]{croft}
Croft R.A.C., Dalton G.B., Esfstathiou G., 1999, \mnras, 305, 547
\bibitem[]{2power}
da \^{A}ngela J., Outram P.J., Shanks T., Boyle B.Bj., Croom S.M., Loaring
N.S., Miller L., and Smith R.Rj., \mnras, 360, 1040

\bibitem[]{croom}
Croom S.M. \etal, 2005, \mnras, 356, 415

\bibitem[]{dalton}
Dalton G.B., Efstathiou G., Maddox S.J., Sutherland W.Wj., 1992, \apj, 390, L1

\bibitem[]{dp83}
Davis, M., and Peebles, P.J.E., 1983, \apj, 267, 465

\bibitem[]{diaferio}
Diaferio A., Borgani S., Moscardini L., Murante G., Dolag K., Springel V.,
Tormen G., Tornatore L., Tozzi P., 2005, \mnras, 356, 1477

\bibitem[]{evrard}
Evrard A.E., MacFarland T., Couchman H.M.P., Colberg J.M., Yoshida N., White
S.D.M. , Jenkins A., Frenk C.S., Pearce F.R., Efstathiou G., Peacock J.A.,
Thomas P.A., 2002, \apj, 573, 7 

\bibitem[]{virgo00}
 Frenk C.S., Colberg J.M., Couchman H.M.P., Efstathiou G., Evrard A.E., 
Jenkins A., MacFarland T.J., Moore B., Peacock J.A., Pearce F.R., 
 Thomas P.A., White S.D.M.,  Yoshida N., astro-ph/0007362

\bibitem[]{forni}
Forni, O., Aghanim, N., 2004, A\&A, 420, 49


\bibitem[]{hamilton}
Hamilton A.J.S., 1992, \apj, 385, L5


\bibitem[]{hansen2002}
Hansen S.H., Pastro S., Semikoz D.V., \apj, 573, L69

\bibitem[]{hansen}
Hansen, S.H., 2004, astro-ph/0410004


\bibitem[]{hawkins}
Hawkins E. \etal, 2003, \mnras, 346, 78

\bibitem[]{hernandez}
Hernández-Monteagudo C. Verde L., Jimenez R., Spergel D.N., 2006 \apj 653, 598

\bibitem[]{hoyle}
Hoyle F., Outram P.J., Shanks T., Boyle B.J., Croom S.M., Smith R.J., 2002,
\mnras 332, 311 

\bibitem[]{hogg}
Hogg D., 1999, astro-ph/9905116

\bibitem[]{haehnelt}
Haehnelt M.G., Tegmark M., 1996, \mnras, 279, 545

\bibitem[]{holder}
Holder G.P., 2004, \apj, 602, 18

\bibitem[]{fof}
Huchra J.P., Geller M.J., 1982, \apj, 257, 423

\bibitem[]{kaiser}
Kaiser N, 1987, \mnras, 1987, 227, 1

\bibitem[]{kashilinsky}
Kashlinsky, A., Atrio-Barandela, F. 2000, \apj, 536, L67

\bibitem[]{knox}
Knox, L., Holder, G., and Church, S.E., 2004, \apj,612, 96

\bibitem[]{krav}
Kravtsov A.V., Church S.E., Hoffman Y., 2002, \apj, 571, 563

\bibitem[]{SO}
Lacey C., Cole S., 1994, \mnras, 271, 676

\bibitem[]{lange}
Lange A. E., Church S. E., Holzapfel W., 1998, American Astronomical Society
Meeting, 30, 908

\bibitem[]{mathiesen}
Mathiesen B.F., Evrard A.E., 2001, \apj, 546, 100

\bibitem[]{matsuto}
Matsubara T., Suto Y., 1996, \apj, 470, L1

\bibitem[]{mcdonald}
McDonald, P., 2003, \apj, 585, 34

\bibitem[]{Mo1996}
Mo H. J., White S. D. M., 1996, \mnras, 282, 347

\bibitem[]{mo1996}
Mo H.J., Jing Y.P., White S.D.M., 1996, \mnras, 282, 1096

\bibitem[]{moscardini}
Moscardini L, Matarrese S., Lucchin E., Rosati P., 2000, \mnras, 316, 283

\bibitem[]{nagai}
Nagai D., Kravtsov A., Kosowsky A., 2003, \apj, 587, 524

\bibitem[]{padilla}
Padilla N.D., Baugh C.M., 2002, \mnras, 329, 431


\bibitem[]{phillipps}
Phillipps S., 1994, \mnras, 269, 1077

\bibitem[]{pointecouteau}
Pointecouteau, E., Giard, M., Barret, D. 1998 A\&A, 336, 44

\bibitem[]{popowski}
Popowski P., Weinberg D., Ryden B., Osmer P., 1998, \apj, 498, 11

\bibitem[]{rephaeli}
Rephaeli Y., 1995ARA\&A, 33, 541

\bibitem[]{seo}
Seo H.-J., Eisenstein D.~J., 2003, \apj, 598, 720

\bibitem[]{sehgal}
Sehgal N., Kosowsky A., Holder G., 2005, \apj, 635, 22

\bibitem[]{springel}
Springel V. \etal, 2005, Nat, 435, 629

\bibitem[]{SZ}
Sunyaev R.A., Zel'dovich Y. B., 1972, Comments Astrophysics and Space Physics,
4, 173 

\bibitem[]{SZ}
Sunyaev R.A., Zel'dovich Y. B., 1980, ARA\& A, 18, 537

\bibitem[]{valeandwhite}
Vale, C. \& White, M., 2006, New Astronomy, 11, 207

\bibitem[]{zahavi2002}
Zehavi I., \etal 2002, \apj, 571, 172

\bibitem[]{zahavi2005}
Zehavi I., \etal 2005, \apj, 630, 1


\end{thebibliography}
\end{document}